# Non-contact electrical detection of intrinsic local charge and internal electric field at nanointerfaces

Duanjun Cai [1], Fuchun Xu [2], Jinchai Li [2], Hangyang Chen [2], and Junyong Kang [2]

[1] Centre of Computational Physics, Physics Department, University of Coimbra, Rua Larga, 3004-516 Coimbra, Portugal

[2] Semiconductor Photonics Research Center, Department of Physics, Xiamen University, Xiamen 361005, China

E-mail: dcai@phys.ntu.edu.tw, jykang@xmu.edu.cn

A nanoscale non-contact electrical measurement has been developed based on Auger electron spectroscopy. This approach used the speciality of Auger electron, which is self-generated and free from external influences, to overcome the technical limitations of conventional measurements. The detections of the intrinsic local charge and internal electric field for nanostructured materials were achieved with resolution below 10 nm. As an example, the electrical properties at the GaN/AlGaN/GaN nanointerfaces were characterized. The concentration of the intrinsic polarization sheet charges embedded in GaN/AlGaN nanointerfacial layers were accurately detected to be -4.4 e/nm$^2$. The mapping of internal electric field across the nanointerface revealed the actual energy band configuration at the early stage of the formation of two-dimensional electron gas.

# 1. Introduction

Characterization of electrical properties is crucial in the exploration of novel nano-materials and nano-devices. Conventional electrical measurements are mainly based on and limited by the construction of circuits [1, 2], wherein the electrodes and external bias voltages have to be introduced. Thus, the contact effects and additional electric field always influence and sometimes alter the electrical properties under investigation [3], preventing us to directly observe the intrinsic properties. Such problems appear to be more significant in low-dimensional systems [4, 5]. Although a lot of effort has been made, present techniques often show their insufficiency in nanoscale measurements [6–8] for nano-structured materials. Therefore, re-describing of electrical quantities at the electronic level would be important for exploring advanced electrical measurements.

From quantum mechanics, it is known that the properties of matter are measurable quantities and can be described by the electronic wave functions. In other words, the electron states contain the full information of the nature of a matter and the local electron states will reflect the specific properties of the local environment. The main challenge for setting up the measurement is to establish the appropriate methods to decipher those specific electron states responsible for different properties. In principle, the electrical properties such as the charge density and electric field are associated with the states and actions of electrons. It can be a new method to develop the electrical measurements by detecting the local electron states.

In this work, our approach is based on the technique of Auger electron spectroscopy (AES). The techniques of intrinsic, high-resolution and non-contact detection of the electrical properties are developed in an electron beam incident AES. This measurement possesses sensitivity and accuracy (2 nm in depth resolution) in the characterizations of the nanoscale materials and structurally complex systems. As an example, the intrinsic local polarization charges and internal electric field at the GaN/AlGaN/GaN nanointerfacial layers are precisely probed and the actual configuration of the energy bands is obtained. The results reveal the basic formation mechanism and the microscopic behavior of different types of local charges embedded at the nanointerfacial layers. The mapping of internal electric field across the nanointerface revealed the actual energy band configuration at the early stage of the formation of two-dimensional electron gas.

## 2. Methods and techniques

The theory and methods of obtaining the electrical signals from the local electron states are proposed based on the knowledge of AES [9-11]. AES stems from the special transition process involving two electrons: the Auger electron is emitted by the energy given to the system from the spontaneous transition of the first electron. The Auger spectrum enables the probing of the electron states of specific atoms in complex systems. The valence band (VB) Auger spectrum (e.g., KVV) can be expressed by the combination of the convolutions of valence electron densities of states (DOS) [12]:

$$\Gamma_{KVV}(E) = a_{pp}D_p(E) \otimes D_p(E) + a_{sp}D_s(E) \otimes D_p(E) + a_{ss}D_s(E) \otimes D_s(E) \quad (1)$$

$$D_l(E) \otimes D_{l'}(E) = \int d\omega D_l(\omega) D_{l'}(E-\omega) \quad (2)$$

where $D_l(E)$ is the partial DOS, $\omega$ is an integral variable, $a_{pp}$, $a_{sp}$ and $a_{ss}$ are the coefficients of the convolution terms. Apparently, the VB Auger spectrum reflects the valence electrons states that are most sensitive to the surrounding environment of an atom and the electron states are correlated and amplified through the convolution integral. This demonstrates that any changes of the local environment, either chemical or physical, will have their direct or indirect signatures in the Auger spectral lineshape. Moreover, because the Auger electrons are generated by spontaneous transitions, which are free from the influence of the primary incident beam, the Auger spectrum can well show the intrinsic properties of an object.

According to the above, at least two parts of the electrical information could have been enfolded in the Auger spectrum, i.e., the electric field and the electron density. Here we will discuss the principles of obtaining the relation between these two electrical quantities and the Auger spectral lineshape. Then, the practical measurement techniques can be established by these principles. As we know, the Auger electron has a fixed kinetic energy. When the additional electric field is applied, the kinetic energy of Auger electrons is changed and consequently, this change will result in the energy shift of Auger peak. Therefore, the Auger spectral shift directly represents the magnitude and direction of the local electric field. This relation will be used to establish the electric field detection by the Auger electron probe. For the latter, the description of electron density in the Auger spectrum will be discussed below. In Eq. (1), one can see that the VB Auger peak consists of the convolution integral of electron DOS's, thereby the peak intensity is associated with the electron density around the corresponding atoms. In order to quantify the relation between the peak height and the electron density, $Al_xGa_{1-x}N$ samples with different Al mole fractions, up to 53 %, were prepared and systematically measured and *ab initio* calculations performed on the same systems [13-15].

Figure 1 shows the N KVV Auger lines for different Al compositions. The three peaks $N_{pp}$, $N_{sp}$ and $N_{ss}$ correspond to the convolution terms of *s*- and *p*-state DOS. One can see that the $N_{pp}$ height appears proportional to the Al mole fraction. This can be explained by the local electron concentration around N atoms as a function of the Al mole fraction. The calculated difference of charge density between the Al-N and Ga-N bonds are shown in the inset of figure 1. It can be seen that the Al-N bond has a stronger polarity than the Ga-N bond has and thus, the higher electron density around N occurs in the Al-N bond. This indicates that the higher composition of Al-N bonds will lead to higher average electron density around N atoms. By using Eq. (1), we find that the higher electron density will give rise to a larger $N_{pp}$ height. Consequently, the quanttive relation between the $N_{pp}$ height and local electron concentration of nitrogen atoms can be determined. Note that the peak height ratio $N_{pp}/N_{sp}$ will be used for our analysis instead of the $N_{pp}$ height. Figure 2 shows this relation between the $N_{pp}/N_{sp}$ ratio and the local electron concentration. Based on above approaches, the measurements can be established for the intrinsic local electric field and local charge concentration.

Electrical measurements were setup on an electron excited AES system, as shown in figure 3. The 5~30 KeV incident electron gun was used for the ionization of the core levels to initiate the Auger process. Emitted electrons are deflected around the electron gun and pass through an aperture towards the back of the cylindrical mirror analyzer (CMA). An axial CMA was employed for the Auger electron collection and the analysis of Auger electron energy spectrum. The Auger peak position was calibrated on clean copper with an accuracy of 0.01 eV. In addition, the system was equipped with a scanning electron microscopy and an $Ar^+$ ion gun. The spatial resolution was better than 10 nm. The ultra high vacuum chamber was kept at a base pressure of $10^{-7}$ Pa or less. Some details of the AES system have also been described in Ref.[10]. In electrical detections, the influences by the external electromagnetic field were eliminated by making the sample well protectively shielded and electrically conductive. Before the electrical measurements, the SEM image was recorded, the chemical compositions were determined, and the elemental mapping was carried out for the identification of the complex structures (surface and heterointerfaces) of the sample. In order to obtain the accurate height of Auger peaks for the determination of ratio $N_{pp}/N_{sp}$, finer energy steps of 0.1 eV and larger signal survey circles (n > 100) were set for the data survey. To probe the internal electric fields, the core-shell Auger spectrum (e.g., Ga LMM peak) was chosen for preventing the sensitive shift of VB spectrum by other physical and/or chemical factors.

## 3. Detection of intrinsic local charges at nanointerfaces

Nitride heterostructures have played a key role in the fabrication of advanced optoelectronic devices and recently rapid progresses have been made in varieties of low-dimensional structures [16-18]. As a strong polar material, the polarization field and charges at the nitride heterointerfaces are very important issues to study [19, 20]. As an example, the widely used GaN/Al$_{0.23}$Ga$_{0.77}$N/GaN heterostructure was prepared by metallorganic chemical vapor deposition [21] and applied for the electrical characterizations. First of all, because of the important phenomenon of heterointerface, compositional diffusion [22, 23], the location and thickness of the interfacial region should carefully determined. For this purpose, the secondary electron imaging and elemental mapping were carried out on the cross-section of the sample by AES, as shown in figure 4a-c. The heterointerfaces can be explicitly distinguished. The Al and Ga elements undergo an inter-diffusion across the heterointerface and a chemically gradient nanointerfacial region forms. More refined examination of the chemical compositions was obtained by the elemental profiling, as shown in figure 4d. Taking into account the effect of the escape depth and the Ar$^+$ ion energy [22], the inter-diffusion length across the heterointerface is determined to be 4.0 nm. This investigation of the inter-diffusion reveals the actual chemical variation in the nanointerfacial region and provides a more accurate picture of the nanointerface than what would have been coarsely understood by assuming a sharp transition profile.

The intrinsic charge concentrations at the GaN/Al$_{0.23}$Ga$_{0.77}$N/GaN nanointerfaces were measured (figure 5). A sheet of negative charges accumulates at the topmost nanointerface below the surface. The spatial distribution of the sheet charges is closely related to the configuration of the diffused nanointerface. These charges congregate mainly in the nanointerfacial region and the maximal concentration appears sharply at the center of the gradient interfacial region. This indicates that the diffused nanointerface plays an important role in accommodating the polar charges. The width and the chemical gradient of the interfacial region are the crucial parameters influencing the charge distribution. By integrating over the entire nanointerfacial range, the total charge concentration covering the region of the interface 1 is -15.8 e/nm$^2$. These charges at this first interface are contributed partially from the deviated negative charges $-\sigma_p$ induced by the internal polarization field and the shifted surface-state electrons $-\sigma_s$. The proportion of these two types of charges will be determined when the charge concentration at the second interface is discussed below. These results indicate that the charge behavior is influenced not only by the polar property of GaN but also by the surface states. The surface-state charges will drift down from the surface and accumulate in the sublayers. This fact can also explain some interesting phenomena observed in nanostructured materials, such as nanorod and nanocluster, where surface effects are significant.

In contrast, another sheet of positive charges accumulates at the second interface with lower concentration of 4.4 e/nm$^2$. It distributes symmetrically across the gradient region and the main part of the charges is confined in the nanointerfacial region. Because this electrical detection is free from interferences by external fields, the charge concentration obtained here directly reveals the behavior of the intrinsic polarization charges $+\sigma_p$. This type of charge has been widely discussed [6, 24, 25], but is rarely detectable by conventional techniques. By the charge-neutrality constraint, one can obtain the corresponding polarization charge concentration at the topmost nanointerface: $-\sigma_p$ = -4.4 e/nm$^2$. This value agrees well with the theoretical prediction of the same order, but with higher spatial accuracy than that by C-V profile measurement [25]. Thus, the ratio of $\sigma_s/\sigma_p$ at the first nanointerface is 2.6. The opposite sign of the polar charges at two interfaces is due to the the arrangement of the GaN and AlGaN heterolayers along the [0001] direction. On the other hand, in the GaN/AlGaN/GaN heterostructure of GaN-based high-power transistor, two-dimensional electron gas (2DEG) has been observed near the second interface [20]. However, the sign of the 2DEG is negative, which is opposite to what we observed at the interface 2. To explain this and understand the mechanism of the formation of these different kinds of charges, investigations of the configuration of electronic energy bands is necessary, which will be discussed below.

## 4. Detection of internal electric fields at nanointerfaces

In general, design of semiconductor heterostructures is the most important method of energy-band engineering. Especially for the strong piezoelectric materials, e.g., GaN and ZnO, the polarization electric fields impose significant bending on the configuration of energy bands. Although theoretical calculations have been applied to the simulation of band configuration of heterostructures [26, 27], direct experimental imaging is still difficult to obtain. In our approach, we attempted to measure the variation of local electric field and thereby, rebuilt the actual energy-band configurations. The longitudinal profile of the local electric field in the GaN/Al$_{0.23}$Ga$_{0.77}$N/GaN across two nanointerfaces was carried out and is shown in figure 6. By the mark-shift technique [10], the Ga LMM peak shift was used to measure the direction and magnitude of the electric field. One can observe in figure 6 that a negative electric potential exists at the first interface whereas a positive electric potential appears at the second interface. Thus, the internal electric field $E$ within AlGaN layers is parallel to the [0001] direction and the average magnitude is about 0.44 V. This internal electric field, in fact, represents the build-in field against the direction of polarization. It is generated by the polarization charges $\pm\sigma_p$ at two nanointerfaces as observed previously. The build-in field and its distribution play an important role in deciding the behavior of carriers and the electronic properties. This point could be clearly seen in the corresponding band configuration.

To get the actual band configuration, the entire flat band of GaN/Al$_{0.23}$Ga$_{0.77}$N/GaN is obtained from the chemical composition profile (figure 4d) by Vegard's law. Then by putting the local electric fields on the flat band, the actual configurations of the conduction band (CB) and VB are rebuilt, as shown in figure 7. The local electric fields introduce additional bending on the energy-band configurations and potential wells form at the nanointerfacial regions. In general, the potential well in CB acts as an electron trap while the VB well as a hole trap (positive charges). In this case, one can see that the CB well appears at the second nanointerface and covers the gradient region. The well width is well consistent with the distribution width of $+\sigma_p$. As shown in figure 7, this CB well traps the free electrons to form the layer of 2DEG. Because the 2DEG comes from the free carriers other than intrinsic valence charges, its negative sign is opposite to the intrinsic $+\sigma_p$. On the other hand, the internal build-in field $E$ leads to the important driving force for the drift and accumulation of the 2DEG. This driving force can be reflected by the difference of electric potential between two nanointerfaces (80 ~ 200 nm), which is about 0.24 eV. The formation of the 2DEG in electronic devices greatly improves the mobility without intentional doping of impurities [20]. These results demonstrate the effect of strong polarization on the different kinds of carriers and explain the forming mechanism and distributing rule of the 2DEG. Since the band configuration can be experimentally mapped by this electrical measurement, detections of the band configuration can be carried out for materials with complex structures. Meanwhile, the band configuration obtained in experiment can be directly compared with the results of theoretical calculations.

## 5. Conclusion

We proposed a method of electrical measurements based on the theory of Auger electronic spectra. The techniques of the non-contact detection of intrinsic local charges and internal electric fields were developed based on AES to achieve high spatial resolution below 10 nm. As an example, the intrinsic polarization charges embedded in the GaN/AlGaN/GaN nanointerface were measured, which was accurately determined to be -4.4 $e/nm^2$. The internal local electric field across two nanointerfaces was detected and thereby the actual band configuration in the early stage of 2DEG formation was depicted. It showed that the width and depth of the potential well at the nanointerfacial regions leads to this accumulation and determines the concentration of the 2DEG therein. Further detections on interesting nano-materials (nanorods and nanoparticles) are in progress.

## Acknowledgement

The authors thank Dr. Apostolos Marinopoulos for helpful discussions. The authors acknowledge the support from the Special Funds for Major State Basic Research Projects and the National Nature Science Foundation of China.

**Figure Captions**

**Figure 1.** (color online) N *KVV* Auger spectra of $Al_xGa_{1-x}N$ in different Al mole fractions. The height of $N_{pp}$ increases with the Al composition increasing. The inset is the subtraction of the charge contour of Al-N bond by Ga-N bond.

**Figure 2.** (color online) Peak height Ratio $N_{pp}/N_{sp}$ as a function of Al mole fraction *x*, fitting with the function of the local electron concentration of $Al_xGa_{1-x}N$.

**Figure 3.** (color online) Schematic of AES electrical measurement.

**Figure 4.** (color online) $GaN/Al_xGa_{1-x}N/GaN$ heterostructures. (a) SEM image of the cross section. (b) and (c) AES elemental mapping of Ga and Al concentrations, respectively, for the nanointerface of (a). (d) Depth profile of elemental compositions of Ga and Al.

**Figure 5.** (color online) Distribution of the local polarization charges at the nanointerfaces of $GaN/Al_{0.23}Ga_{0.77}N/GaN$. The negative and positive charges are localized at different interfaces.

**Figure 6.** (color online) Internal local electric field across the $GaN/Al_{0.23}Ga_{0.77}N/GaN$ nanointerfaces.

**Figure 7.** (color online) Actual band configurations rebuilt by the flat band and local electric field. Dotted lines are the flat bands obtained from figure 3d. *P* and *E* represent the polarization and internal electric field, respectively. The potential well at the second nanointerface traps the 2DEG.

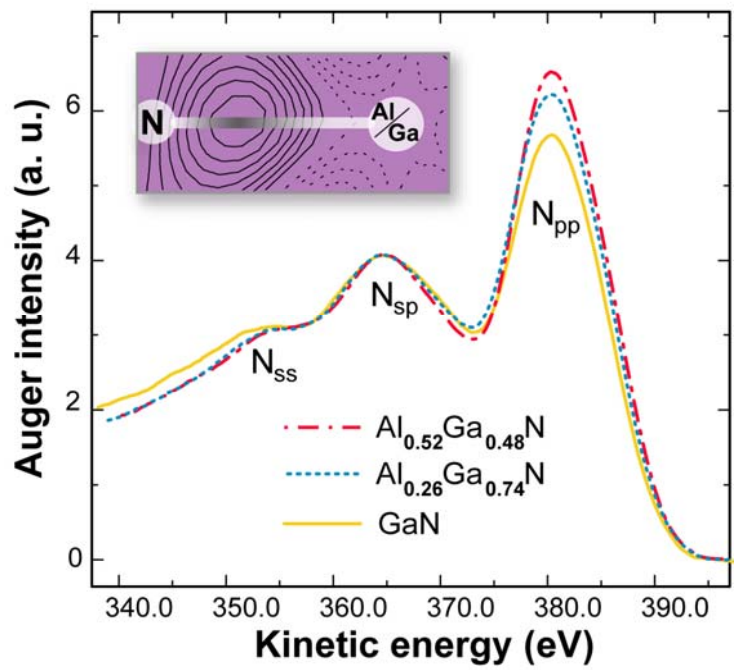

Figure 1. D. Cai et al.

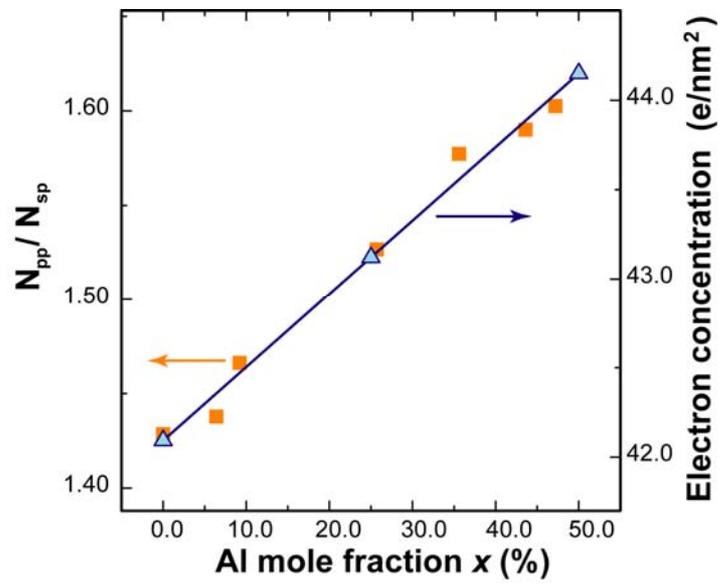

Figure 2. D. Cai et al.

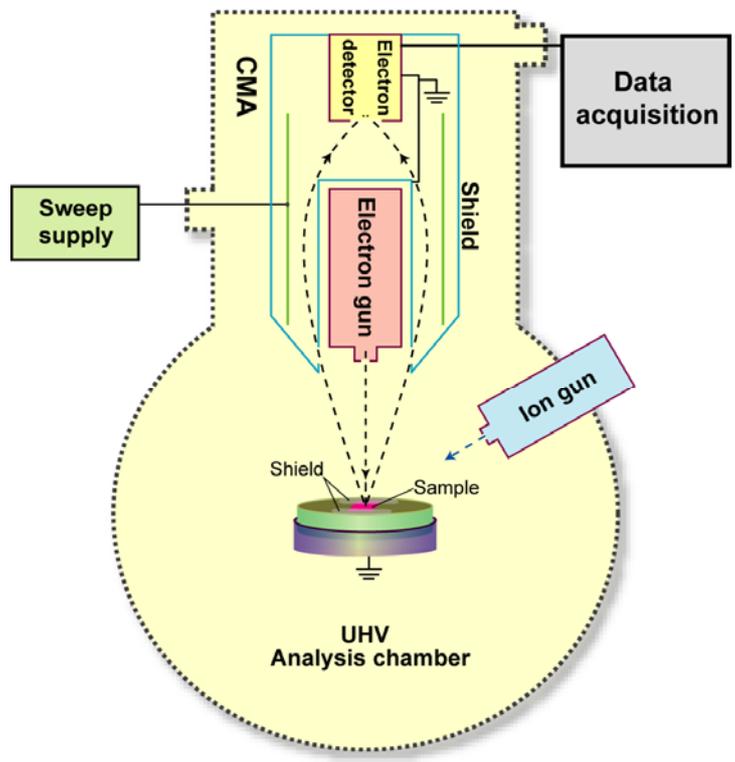

Figure 3. D. Cai et al.

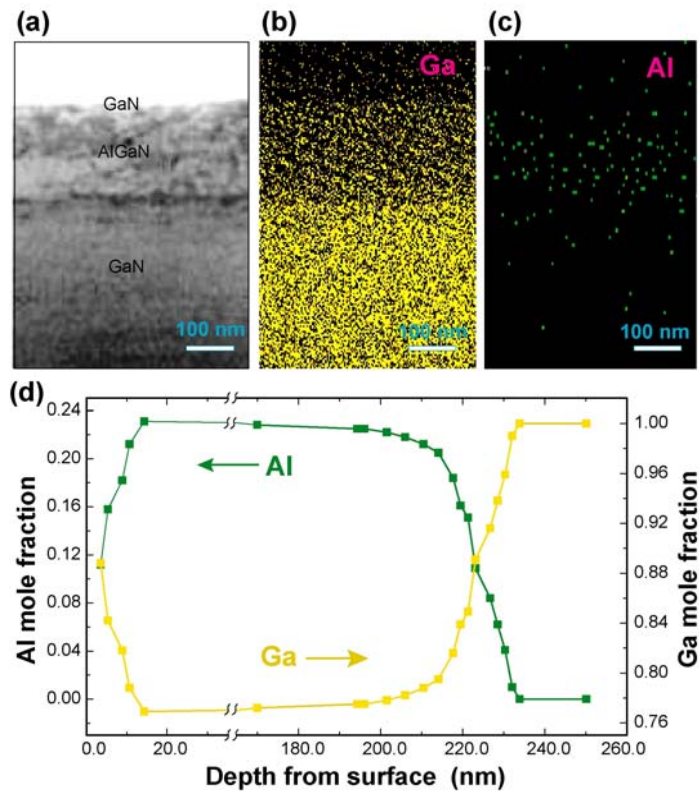

Figure 4. D. Cai et al.

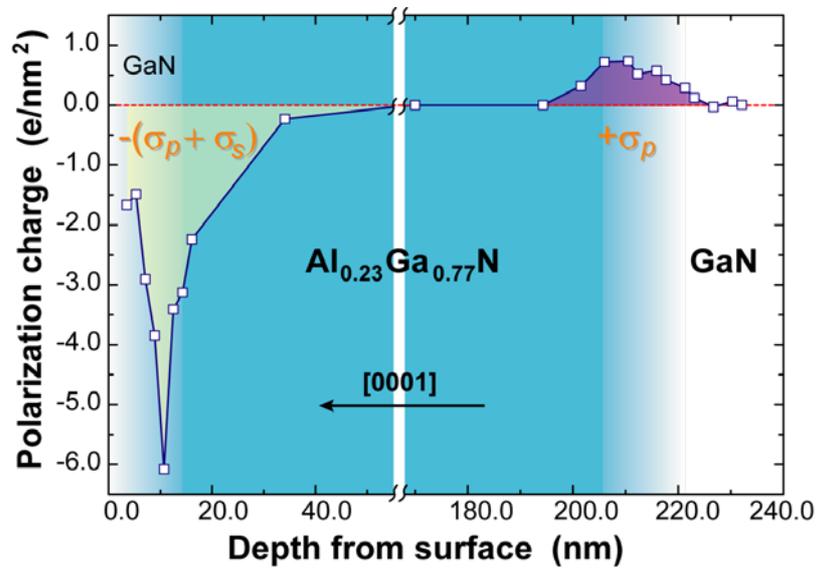

Figure 5. D. Cai et al.

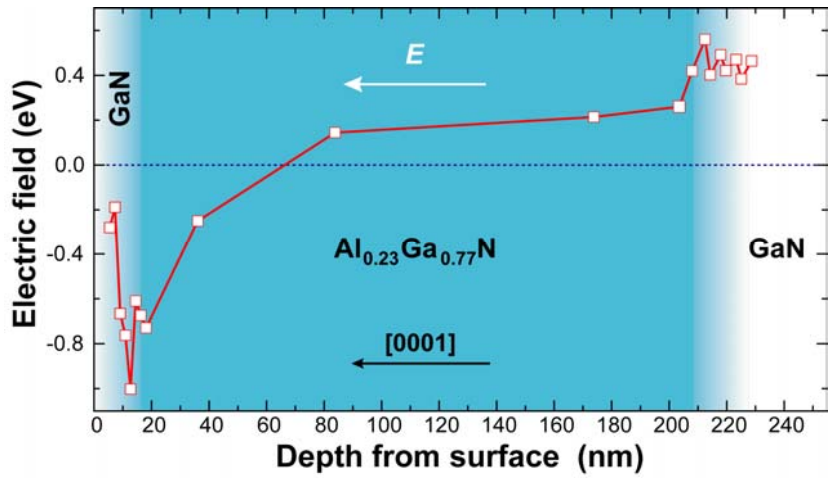

Figure 6. D. Cai et al.

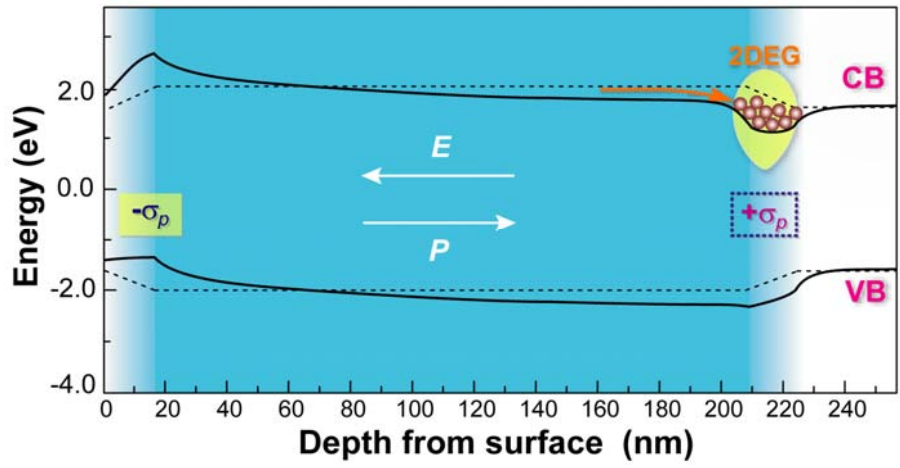

Figure 7. D. Cai et al.